\documentstyle[aps,twocolumn,prb,epsf]{revtex}

\draft
\preprint{Version \today}
\title{Hysteresis effect due to the exchange Coulomb interaction in 
short-period superlattices in tilted magnetic fields}

\author{Andrei Manolescu$^1$ and Vidar Gudmundsson$^2$}

\address{
$^1$Institutul Na\c{t}ional de Fizica Materialelor, C.P. MG-7 
Bucure\c{s}ti-M\u{a}gurele, Rom\^ania,\\
$^2$Science Institute, University of Iceland, Dunhaga 3, IS-107 Reykjavik, 
Iceland}

\begin{document}
%\tighten
\maketitle

\begin{abstract} 

We calculate the ground-state of a two-dimensional electron gas in
a short-period lateral potential in magnetic field, with the Coulomb
electron-electron interaction included in the Hartree-Fock approximation.
For a sufficiently short period the dominant Coulomb effects are
determined by the exchange interaction. We find numerical solutions
of the self-consistent equations that have hysteresis properties when
the magnetic field is tilted and increased, such that the perpendicular
component is always constant.  This behavior is a result of the interplay
of the exchange interaction with the energy dispersion and the
spin splitting.  We suggest that hysteresis effects of this type could
be observable in magneto-transport and magnetization experiments on 
quantum-wire and quantum-dot superlattices.

\end{abstract}


\draft

A well known manifestation of the Coulomb exchange interaction in a
two-dimensional electron gas (2DEG) in a perpendicular magnetic field
is the enhancement of the Zeeman splitting for odd-integer filling
factors,\cite{Ando74:1044} observable in magetotransport experiments
on GaAs systems.  The same mechanism leads to the enhancement of the Landau
gaps for even-integer filling factors, which can be identified in more
recent magnetization measurements.\cite{Meinel99:819}

In the presence of a periodic potential the Landau levels become
periodic Landau bands, and the calculations based on the Hartree-Fock
approximation (HFA) show an enhancement of the energy dispersion of the
bands intersected by the Fermi level.\cite{Manolescu95:1703} Such an effect
has been indirectly observed in the magnetoresistance of short-period
superlattices as an abrupt onset of the spin splitting of the Shubnikov-de
Haas peaks, occurring  only for a sufficiently 
strong magnetic field. \cite{Petit97:225} 
In other words, when the magnetic field increases the
systems makes a first-order phase transition from spin-unpolarized to
spin-polarized states.  This effect has also been discussed in other
forms, for narrow quantum wires, \cite{Kinaret90:11768} and for edge
states.\cite{Dempsey93:3639,Rijkels94:8629}

In the spirit of the HFA, the Coulomb interaction can be split into a
direct and an exchange component.  The direct (classical) interaction
is repulsive (i.\ e.\ positive) and long ranged, while the exchange
(quantum mechanical) part is attractive (i.\ e.\ negative) and short ranged.
The direct component is usually much larger than the exchange one.  In our
system this is decided by the two lengths involved, the superlattice (or
modulation) period $a$, and the magnetic length $\ell=\sqrt{\hbar/eB_0}$
determined by the perpendicular magnetic field $B_0$.  For long periods,
$a\gg\ell$, the screening (direct) effects are strong: the width of the
Landau bands is typically much smaller than the amplitude of the periodic
potential, except when a gap is eventually present at the Fermi level.
\cite{Manolescu97:9707} For periods of the order of $\ell$ the situation
becomes opposite: the screening effect is weak, the exchange interaction
is the dominant Coulomb manifestation, and the energy dispersion of the
Landau bands may exceed the amplitude of the periodic potential if the
latter is small enough.\cite{Manolescu99:5426} 

In a recent paper \cite{Manolescu99:5426} we have studied the numerical
solutions of the Hartree-Fock equations in the presence of short-period
potentials.  After preparing the solution for a fixed potential we
change the potential amplitude by a small amount and find a new,
perturbed solution, and then we change again the amplitude, and repeat
the scheme.  In this way, by increasing and then decreasing the amplitude,
we obtain a hysteretic evolution of the ground state due to the combined
effects of the external potential and of the exchange interaction, on the
energy dispersion of the Landau bands.  In the present paper we consider
a fixed modulation amplitude, but a tilted magnetic field, such that we
include in the problem, self-consistently, the Zeeman splitting of the
Landau bands.  We hereby intend to suggest further experiments that can 
identify strong effects of the Coulomb exchange interaction.  
The material parameters are those for
GaAs: effective mass $m_{\mathrm eff}=0.067 m_e$, dielectric constant
$\kappa=12.4$, bare g-factor $g=-0.44$, and electron concentration
$n_s=2.4 \times 10^{11}$ cm$^{-2}$.

We fix the component of the magnetic field perpendicular to the 2DEG,
$B_0$, which determines our filling factors, while the bare Zeeman
splitting is given by the total field $B=B_0/\cos\phi$, where
$\phi$ is the tilt angle.  We first consider a periodic potential 
varying only along one spatial direction, $V\cos Kx$, where $K=2\pi/a$,
and solve for the eigenstates of the Hamiltonian within the thermodynamic
HFA.  We chose the Landau gauge for the vector potential and we diagonalize
the Hamiltonian in the Landau basis 
$\psi_{nX_0}(x,y)=L_y^{-1/2} e^{-iX_0y/\ell^2} f_n(x-X_0)\mid\sigma\rangle$, 
where $X_0$ is the so-called center coordinate, $L_y$ is the linear
dimension of the 2DEG, $f_n(x-X_0)$ are shifted oscillator 
wave functions, and $\sigma=\pm1$ is the spin projection.

We begin our calculations with $\phi=0$, and find the numerical
HFA-eigenstates by an iterative method, starting from the noninteracting
solution.  Then, we increase $\phi$ and find a new solution starting
from the previous one.  In Fig.\ 1(a) we show a typical energy spectrum,
i.\ e.\ the Landau bands $E_{n\sigma X_0}$, $n=0,1,2,...$, within the first
Brillouin zone, for a small tilt angle, $\phi < \phi_1$.  Here $B_0=4.1$
T, and the parameters of the external potential are $a=40$ nm and $V=9$
meV.

In a simplified view, the exchange interaction contributes
with a negative amount of energy to the occupied states, which
enhances the energy dispersion in the vicinity of the Fermi
level.\cite{Manolescu95:1703,Manolescu99:5426} The classical Hartree
(positive) energy is small in our case, but it would increase with
increasing modulation period and would rapidly flatten the energy
bands.  Also, the spin splitting is almost suppressed for $\phi < \phi_1$.
However, for a sufficiently high field, when $\phi=\phi_1$, the difference
in population of the spin-up and spin-down bands exceeds a critical
value, and the spin gap is abruptly amplified by the exchange energy.
The spin-up states become self-consistently more populated and lower
in energy.  The energy spectrum becomes like in Fig.\ 1(b), and keeps
this structure when $\phi$ further increases.  Then, we decrease $\phi$
step by step. For low temperatures we find for $\phi=0$ the solution
with large spin gap, similar to Fig.\ 1(b), while for higher temperatures
we may find a transition to the spin-unpolarized state, Fig.\ 1(a),
at $\phi_2 < \phi_1$.

We show in Fig.\ 2 the spin polarization,
$(n_{\uparrow}-n_{\downarrow})/(n_{\uparrow}+n_{\downarrow})$, for two
temperatures, when $\phi_2>0$.  We consider here the temperature 
only as an effective parameter, that may also include includes the 
effects of a certain disorder, inherent in any real system.  Clearly, 
in the presence of disorder similar results will appear for lower 
temperatures.

We have explicitly included disorder in a transport calculation,
by assuming Gaussian spectral functions, $\rho_{n\sigma}(E)=
(\Gamma\sqrt{\pi/2})^{-1}\exp[-2(E-E_{n\sigma})^2/\Gamma^2]$,
where $\Gamma$ is the Landau level broadening.  We have
calculated the conductivity tensor $\sigma_{\alpha\beta}$,
$\alpha,\beta=x,y$, using the standard Kubo formalism for the
modulated 2DEG.  \cite{Zhang90:12850,Manolescu97:9707} In Fig.\
3 we show the hysteresis loops for the longitudinal resistivities
$\rho_{xx,yy}=\sigma_{yy,xx}/(\sigma_{xx}\sigma_{yy}+\sigma_{xy}^2)$.
In our regime $\sigma_{xy}^2\gg\sigma_{xx}\sigma_{yy}$, such that
$\rho_{xx,yy}$ are in fact proportional to $\sigma_{yy,xx}$.  Also,
the conductivity in the $y$ direction is dominated by the quasi-free
net electron motion along the equipotential lines of the  modulation,
with group velocity $<v_y>=-(eB_0)^{-1}dE_{nX_0}/dX_0$, known as band
conductivity,\cite{Aizin84:1469} 
while the conductivity in the $x$ direction is related to
inter-band scattering processes (scattering conductivity).  The band
and the scattering conductivities are inverse and respectively direct
proportional to a power of the density of states at the Fermi level
(DOSF).\cite{Zhang90:12850,Manolescu97:9707}  In the transition between
spin-unpolarized and spin-polarized states the Fermi level touches
the minima of the band $E_{1\downarrow}$, where DOSF has a van Hove
singularity. Therefore, for that situation the band conductivity, and
thus $\rho_{xx}$, have a minimum, whereas the scattering conductivity,
and thus $\rho_{yy}$, have a maximum, see Fig.\ 3.

Similar hysteresis effects can be found in a 2DEG that is modulated
in two perpendicular spatial directions.  
\cite{Manolescu99:5426,Gudmundsson95:16744} 
However, in this case the picture is further
complicated by the presence of the Hofstadter\cite{Hofstadter76:2239}
gaps and their interplay with the spin gaps.  Then, since the dispersion
of the Landau bands is essential for the hysteresis, another complication
with the two-dimensional potential is that for an asymmetric unit cell
the behavior of the system may not be the same when the magnetic field is
tilted towards the $x$ or towards the $y$ axis of the plane, reflecting
the anisotropy of the Brillouin zones.  These details are not addressed in 
this paper.

In principle, the effects discussed in the present paper should also
occur in narrow quantum wires or dots, and not necessarily only in
periodic systems, as long as Landau bands with both flat and steep
regions are generated, as in Fig. \ 1.  In wide wires or dots
the electrostatic screening is expected to dominate the exchange
interaction, just like in long-period superlattices, and thus the
Landau bands are smooth, except at the edges. 
Rijkels and Bauer \cite{Rijkels94:8629}
have also predicted hysteresis effects in the edge channels of quantum
wires when a small chemical-potential difference between the spin-up and
spin-down channels can be controlled.  To our knowledge an experimental
confirmation has not been reported yet.  Instead, several groups have
build short-period superlattices for transport and other experiments,
\cite{Petit97:225,Nakamura98:944,Schlosser96:683} which can also be used
to check our predictions.

In conclusion, we have found a hysteresis property of the numerical
solution of the thermodynamic HFA, with the physical origin in the
exchange effects of the Coulomb interaction in the quantum Hall regime,
in the presence of a short-period potential, when the Zeeman splitting
is changed by tilting the magnetic field with respect to the 2DEG.
We suggest that such effects could be observed e.\ g.\ in magnetization
or magnetotransport measurements.

A.\ M.\  was supported by a NATO fellowship at the Science Institute,
University of Iceland.  The research was partly supported by
the Icelandic Natural Science Foundation, and the University of Iceland
Research Fund.

%\newpage

\vspace{-2.2cm}
\begin{figure}
\epsfxsize 13.5cm
\begin{center}
      \epsffile{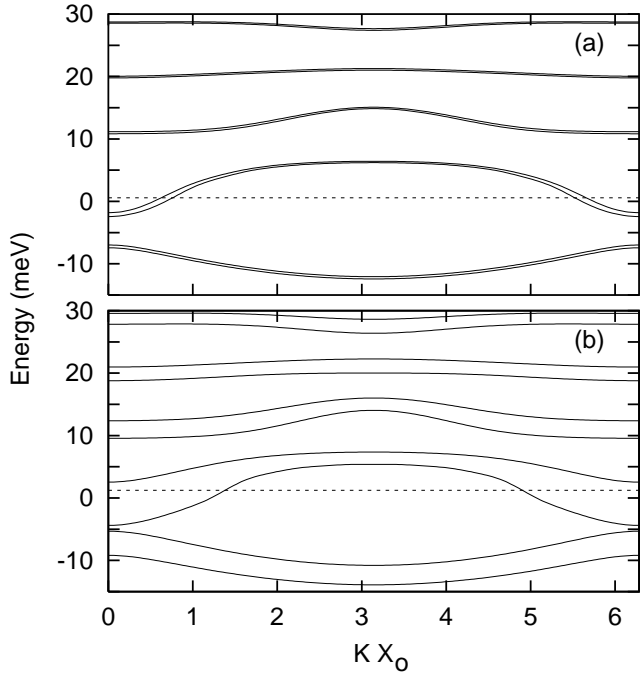}
\end{center}
\vspace{-7cm}
\caption{Two energy spectra: (a) $\cos\phi=0$, (b) $\cos\phi=1/6$.
         The dashed lines show the Fermi level.} 
\end{figure}
\vspace{2cm}
\begin{figure}
\epsfxsize 9cm
\begin{center}
      \epsffile{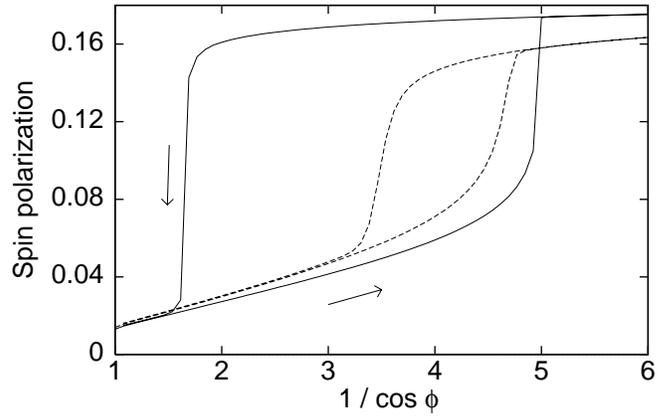}
\end{center}
\caption{Hysteresis loops for the spin polarization for $T=5$~K, 
         with solid line, and for $T=7$ K, with dashed line.} 
\end{figure}
\begin{figure}
\epsfxsize 10cm
\begin{center}
      \epsffile{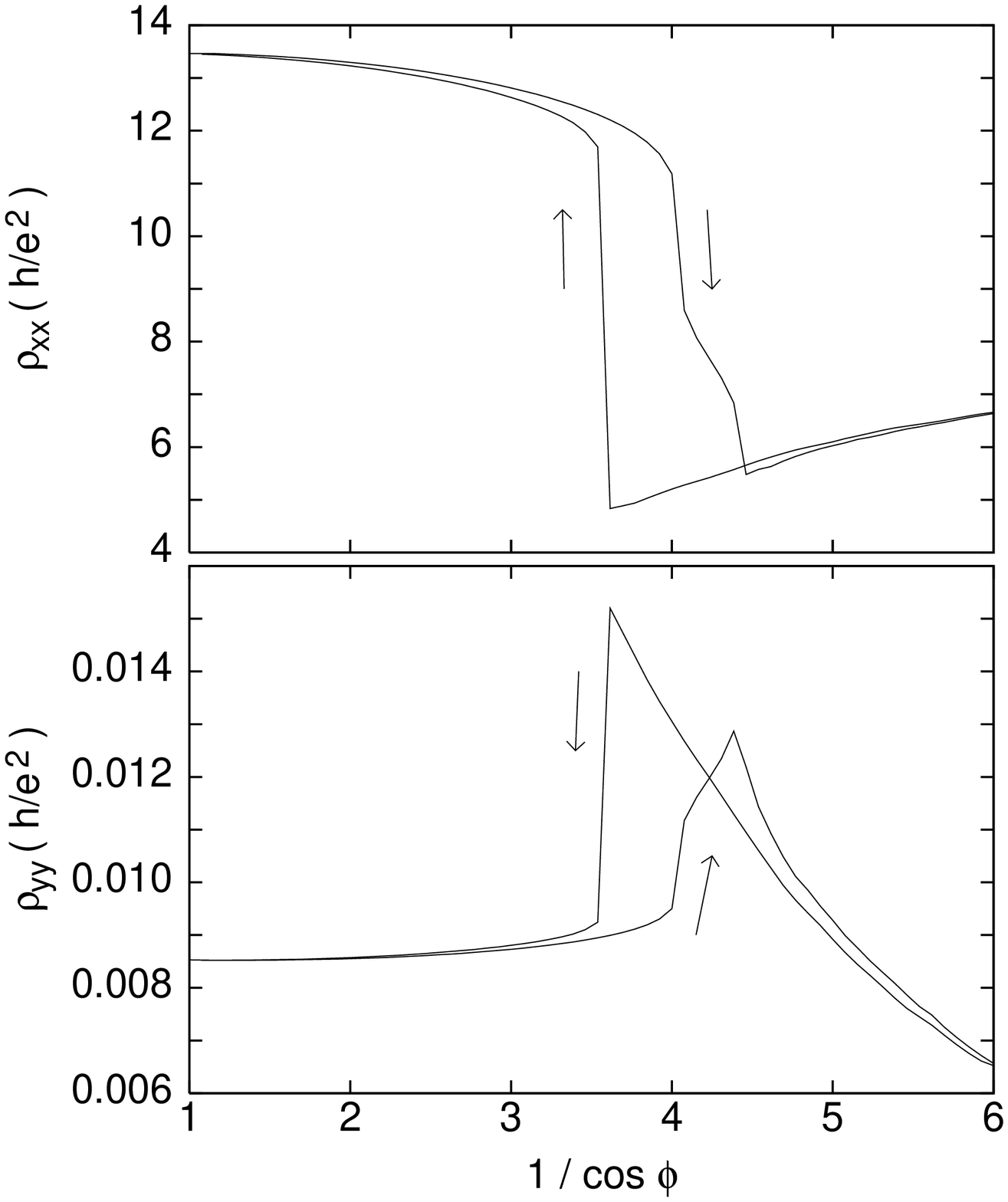}
\end{center}
\vspace{-3cm}
\caption{Hysteresis loops for the resistivities $\rho_{xx}$ and $\rho_{yy}$.
         $T=2$ K, $\Gamma=1$ meV.}
\end{figure}

\begin{thebibliography}{}

\bibitem{Ando74:1044}
T. Ando and Y. Uemura, 
J. Phys. Soc. Jpn. {\bf 37}, 1044 (1974).

\bibitem{Meinel99:819}
I. Meinel, T. Hengstmann, D. Grundler, and D. Heitmann, 
Phys. Rev. Lett.  {\bf 82},  819  (1999).

\bibitem{Manolescu95:1703}
A. Manolescu and R. R. Gerhardts, 
Phys. Rev. B {\bf 51}, 1703 (1995).

\bibitem{Petit97:225}
F. Petit, L. Sfaxi, F. Lelarge, A. Cavanna, M. Hayne, and B. Etienne, 
Europhys. Lett. {\bf 38}, 225 (1997).

\bibitem{Kinaret90:11768}
J. M. Kinaret and P. A. Lee,
Phys. Rev. B {\bf 42}, 11768 (1990).

\bibitem{Dempsey93:3639}
J. Dempsey, B. Y. Gelfand, and B. I. Halperin,
Phys. Rev. Lett. {\bf 70}, 3639 (1993).

\bibitem{Rijkels94:8629}
L. Rijkels and G. W. Bauer, Phys. Rev. B {\bf 50}, 8629 (1994).

\bibitem{Manolescu97:9707}
A. Manolescu and R. R. Gerhardts, 
Phys. Rev. B {\bf 56}, 9707 (1997).

\bibitem{Manolescu99:5426}
A. Manolescu and V. Gudmundsson,
Phys. Rev. B {\bf 59}, 5426 (1999).

\bibitem{Zhang90:12850}
C. Zhang and R. R. Gerhardts
Phys. Rev. B {\bf 41}, 12850 (1990).

\bibitem{Aizin84:1469} 
G. R. Aizin and V. A. Volkov,
Zh. Eksp. Teor. Fiz. {\bf 87}, 1469 (1984)
[Sov. Phys. JETP {\bf 60}, 844 (1984)].

\bibitem{Gudmundsson95:16744}
V. Gudmundsson and R. R. Gerhardts,
Phys. Rev. B {\bf 52}, 16744 (1995).

\bibitem{Hofstadter76:2239}
R. D. Hofstadter,
Phys. Rev. B {\bf 14}, 2239 (1976).

\bibitem{Nakamura98:944}
Y. Nakamura, T. Inoshita, and H. Sakaki, 
Physica E {\bf 2}, 944 (1998).

\bibitem{Schlosser96:683}
T. Schl\"osser, K. Ensslin, J. P. Kotthaus, and M. Holland,
Europhys. Lett. {\bf 33}, 683 (1996).

\end{thebibliography}
\end{document}